\documentclass{article}
\usepackage[preprint]{spconf}
\usepackage{amsmath,graphicx,hyperref}
\usepackage{amssymb}
\usepackage{algorithm} 
\usepackage[noend]{algpseudocode}

\usepackage{tikz}

\usetikzlibrary{shapes.geometric, arrows.meta, positioning, calc}
\usepackage{xcolor}

\usetikzlibrary{external}
\toappear{\parbox{\textwidth}{\centering\footnotesize © 2025 IEEE. This work has been submitted to the IEEE for possible publication. Copyright may be transferred without notice, after which this version may no longer be accessible. Personal use of this material is permitted. Permission from IEEE must be obtained for all other uses, in any current or future media, including reprinting/republishing this material for advertising or promotional purposes, creating new collective works, for resale or redistribution to servers or lists, or reuse of any copyrighted component of this work in other works.}}

\title{Similarity-Guided Diffusion for Long-Gap Music Inpainting}
\name{Sean Turland \quad Eloi Moliner \quad Vesa Välimäki }
\address{Acoustics Lab, Dept.~Information and Communications Engineering, Aalto University, Espoo, Finland}
\usepackage[nolist]{acronym}
\acrodef{stft}[STFT]{short-time Fourier transform}
\acrodef{snr}[SNR]{signal-to-noise ratio}
\acrodef{ode}[ODE]{ordinary differential equation}
\acrodef{dps}[DPS]{diffusion posterior sampling}
\acrodef{cqt}[C$Q$T]{Constant-$Q$ Transform}

\begin{document}
\ninept
\maketitle
\begin{abstract}
Music inpainting aims to reconstruct missing segments of a corrupted recording. While diffusion-based generative models improve reconstruction for medium-length gaps, they often struggle to preserve musical plausibility over multi-second gaps.
We introduce Similarity-Guided Diffusion Posterior Sampling (SimDPS), a hybrid method that combines diffusion-based inference with similarity search.
Candidate segments are first retrieved from a corpus based on contextual similarity, then incorporated into a modified likelihood that guides the diffusion process toward contextually consistent reconstructions.
Subjective evaluation on piano music inpainting with 2-s gaps shows that the proposed SimDPS method enhances perceptual plausibility compared to unguided diffusion and frequently outperforms similarity search alone when moderately similar candidates are available. These results demonstrate the potential of a hybrid similarity approach for diffusion-based audio enhancement with long gaps.
\end{abstract}
\section{Introduction}
\label{sec:intro}

Audio inpainting \cite{adler_audio_2012} attempts to reconstruct missing segments of a signal by replacing them with perceptually plausible content. 
It has a wide range of applications, including packet loss concealment in data transmission \cite{sanneck_new_1996}, restoration of historical recordings from degraded physical media \cite{moliner_diffusion-based_2024}, and correction of errors in post-production audio editing \cite{levy_controllable_2023}.The requirements of these applications vary significantly with respect to the duration of the gaps to be reconstructed: while packet loss concealment typically involves gaps of a few milliseconds, restoration and editing scenarios may require filling much longer segments. This paper tackles the issue of audio inpainting using generative modelling techniques, proposing a solution for longer gaps than previously.

Traditional audio inpainting methods, such as autoregressive modeling \cite{janssen_adaptive_1986,esquef_interpolation_2003,kauppinen_audio_2002} and approaches based on time–frequency sparsity \cite{mokry_introducing_2019,taubock2020dictionary},
rely solely on information immediately surrounding the lost signal segment.
They assume that the audio waveform can be modeled as approximately stationary over short time spans, meaning its spectral characteristics remain predictable from recent past samples. 
These methods are effective for very short gaps, typically below 10\,ms, but their performance deteriorates rapidly for longer durations, as the assumption of local stationarity no longer holds.

Acknowledging the repetitive nature of music signals, Perraudin et al.~\cite{perraudin_inpainting_2018} identified an inpainting solution that represents the entire signal as a graph of \ac{stft} frames. A similarity search optimisation was used to find the best inpainting candidate from other portions of the audio signal. This produced perceptually coherent inpainting solutions of multiple seconds, but performed poorly on cases where no such replacement existed, and can struggle to ensure smooth, seamless transitions.

To address these shortcomings, novel audio inpainting techniques have applied methods from the field of machine learning, specifically probabilistic deep learning. One class of probabilistic deep learning methods, diffusion-based generative models, demonstrated excellent abilities in generating high-quality signals across a wide range of domains \cite{sohl-dickstein_deep_2015, ho_denoising_2020}. The application of diffusion–based posterior sampling to various inverse problems in audio signal processing has been studied
widely 
in the literature \cite{lemercier_diffusion_2024}, and their specific use in audio inpainting demonstrates quality on long corruption lengths beyond 300\,ms \cite{moliner_diffusion-based_2024}. Despite their flexibility, solutions for longer inpainting corruptions of multiple seconds often fail to meet listeners’ perceptual expectations on semantic features, such as melody, loudness or rhythm \cite{moliner_diffusion-based_2024}. 

Using an auxiliary solution to guide the generative output of diffusion models has led to significant results in a variety of domains, including molecular biology \cite{watson2023novo} and image generation \cite{meng_sdedit_2022}. Although previous research into audio signal processing has conditioned the inference procedure on text \cite{novack_ditto_2024} and MIDI \cite{maman_multi-aspect_2025}, 
the combination of similarity search techniques with diffusion posterior guidance methods remains unexplored.

This paper proposes Similarity-Guided Diffusion Posterior Sampling (SimDPS), a hybrid approach that integrates generative diffusion models with guidance from contextually similar audio segments. The method first retrieves an auxiliary segment from a corpus by minimising a cost function that measures similarity to the gap context. This segment is then used to guide the diffusion process, enabling the model to generate a reconstruction that is both perceptually plausible and consistent with the surrounding audio.

The paper is organised as follows. Sec.~2 provides background on audio inpainting and diffusion models. Sec.~3 presents our proposed similarity-guided diffusion framework. Sec.~4 details the experiment specifics. Sec.~5 reports a subjective evaluation on long-gap piano inpainting, and Sec.~6 concludes the paper.

\section{Background}


Audio inpainting can be formulated as a linear inverse problem:
\begin{equation}
    \mathbf{y} = \mathbf{M}\mathbf{x} + \boldsymbol{\varepsilon}_y,
    \label{eq:inverse}
\end{equation}
such that the observed signal $\mathbf{y} \in \mathbb{R}^n$ is the result of transforming the original signal $\mathbf{x} \in \mathbb{R}^n $ with a \textit{mask} matrix $\mathbf{M} \in \mathbb{R}^{n\times n}$
, whilst $\boldsymbol{\varepsilon}_y \sim \mathcal{N}(\mathbf{0}, \sigma^2_y \mathbf{I})$ accounts for additional noise or measurement errors.
The aim of
audio inpainting
is to reconstruct $\mathbf{x}$.

The \textit{mask} matrix $\mathbf{M}$ is diagonal and binary, with elements defined as $\mathbf{M}_{ij} = m_i \delta_{ij}$,
where $\delta_{ij}$ is the Kronecker delta and $m_i \in \{0, 1\}$.
We focus on compact gaps, restricting the mask to a single missing interval $[t_s, t_e]$ of length $L = t_e - t_s$, with
$m_i = 0$ for $t_s \leq i \leq t_e$ and $m_i = 1$ otherwise.

The masking operator $\mathbf{M}$ is inherently non-injective, with a non-trivial null space
$\mathrm{Ker}(\mathbf{M}) = \{ \mathbf{z} \in \mathbb{R}^n \,:\, \mathbf{M}\mathbf{z} = \mathbf{0} \}$,
which prevents the existence of an exact inverse.
Moreover, since $\mathbf{M}$ projects onto the observed indices, its complement $(\mathbf{I}-\mathbf{M})$ acts as the orthogonal projector onto $\mathrm{Ker}(\mathbf{M})$,
so that projecting $\mathbf{y}$ into the null-space
yields zero: $(\mathbf{I} - \mathbf{M})\mathbf{y} = \mathbf{0}$.
As such, the inpainting problem is ill-posed, and reconstruction methods must rely on prior information or structural assumptions to recover perceptually consistent solutions.

\subsection{Diffusion Probabilistic Models}

To address the ill-posed nature of audio inpainting, one can use a data-driven prior that models the distribution of clean signals, $p_x$, which is only accessible through a finite dataset of examples \cite{moliner_diffusion-based_2024}. In this work, we focus on diffusion probabilistic models, a class of generative models capable of sampling from complex distributions which have gained significant research attention in recent years \cite{song_score-based_2021, karras2022elucidating}.
They define an iterative mapping, controlled by a time variable $\tau$, that transforms samples from a tractable prior distribution $p_T$ at $\tau = T$ into samples from the data distribution 
$p_x$ at diffusion time $\tau=0$.
The prior distribution is typically chosen as an isotropic Gaussian $p_T$ = $\mathcal{N}(\mathbf{0}, \sigma^2(T) \mathbf{I})$, which is easy to sample from.

Following the design choices from \cite{karras2022elucidating}, this mapping can be described by the following \textit{Probability-Flow Ordinary Differential Equation}, 
which defines a deterministic, bijective transformation:
\begin{equation}
    d\mathbf{x} = - \dot \sigma(\tau)\sigma(\tau) \  \nabla_{\mathbf{x}} \ \text{log} \ p_{\tau}(\textbf{x}; \sigma(\tau)) \ d\tau \ .
\label{eq:probflow}
\end{equation}
Here, $ \dot\sigma(\tau)$ denotes the derivative of $\sigma(\tau)$ with respect to $\tau$.
Diffusion models approximate the intractable score function with a neural network
$s_\theta(\mathbf{x}_\tau, \tau) \approx \nabla_{\mathbf{x}} \ \text{log}  \, p_{\tau}(\textbf{x}; \sigma(\tau))$, which is trained with a denoising score matching objective \cite{vincent2011connection}.
Using an ODE solver, such as the Euler-Heun method \cite{karras2022elucidating}, we can initialise $\mathbf{x}_T \sim p_T$ and solve (\ref{eq:probflow}) to generate new samples from the prior distribution, $p_x$.

An important property of the score function under a Gaussian corruption process is that it can be expressed in terms of the minimum mean squared error (MMSE) estimator of the clean sample
 $\mathbf{x}_0$, given the noise-corrupted sample, $\mathbf{x}_\tau$. By Tweedie's formula \cite{efron2011tweedie},
\begin{equation}
    \nabla_{\mathbf{x}} \ \text{log} \ p_\tau(\textbf{x}; \sigma(\tau)) = \frac{\mathbb{E}[\mathbf{x}_0 | \mathbf{x}_\tau; \sigma(\tau)] - \mathbf{x}_\tau}{\sigma^2(t)} \ 
    .
    \label{eq:tweedie}
\end{equation}
This connection shows that the score network can be interpreted as a Gaussian denoiser:
$\hat{\mathbf{x}}_0(\mathbf{x}_\tau)= \mathbf{x}_\tau+ \sigma^2(\tau) s_\theta(\mathbf{x}_\tau, \tau) \approx 
\mathbb{E}[\mathbf{x}_0 | \mathbf{x}_\tau; \sigma(\tau)]$.
We adopt the design choices recommended in \cite{karras2022elucidating}, including the linear time parameterization $\sigma(\tau) = \tau$.

\subsection{Diffusion Posterior Sampling}

To enable control over specific attributes of generated samples—such as in audio inpainting—several studies \cite{chung2023diffusion, moliner2023solving} explore sampling from the conditional posterior $p(\mathbf{x}|\mathbf{y})$ by replacing the prior score in (\ref{eq:probflow}) with the posterior score.
Applying Bayes' rule,
the posterior score decomposes as  
\begin{equation}
\underbrace{\nabla_{\mathbf{x}_\tau} \log p_\tau(\mathbf{x}_\tau | \mathbf{y})}_{\text{Posterior Score}} = \underbrace{\nabla_{\mathbf{x}_\tau} \log p_\tau(\mathbf{x}_\tau)}_{\text{Prior Score}} + \underbrace{\nabla_{\mathbf{x}_\tau} \log p_\tau(\mathbf{y} | \mathbf{x}_\tau)}_{\text{Likelihood Score}} \ ,
\label{eq:bayes_score_decomp}
\end{equation}
where the prior score is approximated with the model $s_\theta(\mathbf{x}_\tau, \tau)$.

Following \cite{chung_diffusion_2024, moliner_diffusion-based_2024}, we approximate the $\tau$-dependent likelihood as a Gaussian distribution centered at the denoised estimate
obtained from the score model: $p_\tau(\mathbf{y}| \mathbf{x}_\tau) \approx \mathcal{N}(\hat{\mathbf{x}}_0(\mathbf{x}_\tau); \sigma^2_y \mathbf{I})$.
This leads to the likelihood score approximation:

\begin{equation}
\nabla_{\mathbf{x}_\tau} \log p_\tau(\mathbf{y} | \mathbf{x}_\tau) \approx -\nabla_{\mathbf{x}_\tau} \frac{1}{\sigma_y^2} \left\| \mathbf{y} - \mathbf{M}\hat{\mathbf{x}}_0(\mathbf{x}_\tau) \right\|_2^2 \ ,
\label{eq:likelihood_gradient}
\end{equation}
where the gradient operator $\nabla_{\mathbf{x}_\tau}$ is implemented via automatic differentiation and therefore propagates through the score model $s_\theta(\mathbf{x}_\tau, \tau)$. 
Since the score model is a neural network trained to exploit temporal dependencies in audio waveforms, the backpropagated gradient is not confined to the range of $\mathbf{M}$, but also produces nonzero components in $\mathrm{Ker}(\mathbf{M})$. 
Consequently, the likelihood score provides guidance not only on the observed samples but also indirectly on the missing region through correlations learned during training. In combination with the prior score, it steers the sampling trajectory towards solutions consistent with the observations $\mathbf{y}$.

\section{Similarity-Guided Diffusion}

\begin{figure}
    \centering
    \resizebox{\columnwidth}{!}{%
    \includegraphics[width=1.0\textwidth]{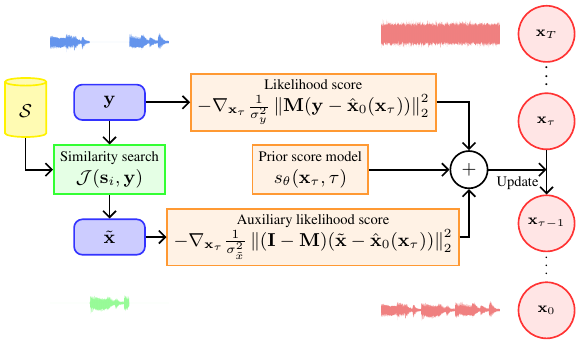}
    }
    \vspace{-10pt}
    \caption{Diagram of the proposed SimDPS framework.}
    \label{fig:diagram}
    \label{fig:placeholder}
\end{figure}

Whilst successful for short- to medium-length gaps \cite{moliner_diffusion-based_2024}, the posterior sampling approach is less effective for longer gaps ($>$300 ms). The likelihood score (\ref{eq:likelihood_gradient}) constrains the generated signal only in the observed regions, leaving the missing portion—lying in $\mathrm{Ker}(\mathbf{M})$—  weakly conditioned.
We observed that, since the diffusion prior encodes local waveform statistics without explicit musical knowledge, reconstructions can diverge from expected musical content despite being locally smooth and acoustically plausible.

To address this, we introduce an auxiliary 
signal $\tilde{\mathbf{x}} \in \mathbb{R}^n$ serving as a first  approximation of the unknown $\mathbf{x}_0$
and containing nonzero components in $\mathrm{Ker}(\mathbf{M})$.
As detailed in Sec.~\ref{sec:sim}, $\tilde{\mathbf{x}}$ is obtained from a corpus of audio signals via a similarity search procedure that uses the observed context $\mathbf{y}$ as reference. 
The retrieved segment is then integrated into a modified likelihood described in Sec.~\ref{sec:unilik}, which is used for guiding the reverse diffusion process through a modified posterior 
An overview of the proposed SimDPS framework, including its posterior sampling components, is shown in Fig.~\ref{fig:diagram}.

\subsection{Modified Likelihood for Similarity-Guided Inpainting} \label{sec:unilik}

We model the relation between the auxiliary signal $\tilde{\mathbf{x}}$ and the ground truth $\mathbf{x}_0$ in the null-space of $\mathbf{M}$ as
\begin{equation} \label{eq:approx}
   (\mathbf{I} - \mathbf{M})  \tilde{\mathbf{x}} =    (\mathbf{I} - \mathbf{M}) (  \mathbf{x} + \boldsymbol{\varepsilon}_{\tilde{x}}), \quad \boldsymbol{\varepsilon}_{\tilde{x}} \sim \mathcal{N}(\mathbf{0}, \sigma_{\tilde{x}}^2 \mathbf{I}),
\end{equation}
where $\boldsymbol{\varepsilon}_{\tilde{x}}$ models the uncertainty of the auxilary variable.
A synthetic measurement is then constructed by
\begin{equation}
    \mathbf{\tilde{y}} = \mathbf{M} \mathbf{y} + (\mathbf{I} - \mathbf{M})\tilde{\mathbf{x}},
    \label{eq:synthetic_measurement}
\end{equation}
combining the observed measurements $\mathbf{y}$ with the auxiliary estimate $\tilde{\mathbf{x}}$ projected into $\mathrm{Ker}(\mathbf{M})$.
Substituting \eqref{eq:inverse} and \eqref{eq:approx} gives
$\mathbf{\tilde{y}} = \mathbf{x} + \boldsymbol{\varepsilon}$,
where $\boldsymbol{\varepsilon} \sim \mathcal{N}(\mathbf{0}, \boldsymbol{\Sigma})$ represents a modified, non-isotropic Gaussian capturing both measurement and auxiliary uncertainties. Its covariance $\boldsymbol{\Sigma}$ is diagonal, with sample-wise variances $\Sigma{ii} = m_i \sigma_y^2 + (1 - m_i) \sigma_{\tilde{x}}^2$.
Owing to the linearity of $\mathbf{M}$, the synthetic log-likelihood decomposes as the sum of two independent terms:
$\log p(\mathbf{\tilde{y}} | \mathbf{x}) = \log p(\mathbf{y} | \mathbf{x}) + \log p(\tilde{\mathbf{x}} | \mathbf{x})$.

The modified likelihood integrates directly into \eqref{eq:bayes_score_decomp}. Using the denoised estimate $\hat{\mathbf{x}}_0(\mathbf{x}\tau)$, the modified likelihood score can be approximated analogously to \eqref{eq:likelihood_gradient} as
\begin{equation}
\begin{split}
    \nabla_{\mathbf{x}_\tau} \log p_\tau(\mathbf{\tilde{y}} | \mathbf{x}_\tau) &\approx{} -\nabla_{\mathbf{x}_\tau}  \frac{1}{\sigma_y^2}\left\| \mathbf{M}(\mathbf{y} - \hat{\mathbf{x}}_0(\mathbf{x}_\tau)) \right\|_2^2 \\
    & \left. - \nabla_{\mathbf{x}_\tau}  \frac{1}{\sigma^2_{\tilde{x}}}\left\| (\mathbf{I}-\mathbf{M})(\tilde{\mathbf{x}} - \hat{\mathbf{x}}_0(\mathbf{x}_\tau)) \right\|_2^2 .\right .
\end{split}
\label{eq:unified_gradient}
\end{equation}
This provides a modified guidance mechanism, where the influence of the observed measurement $\mathbf{y}$ and the auxiliary $\tilde{\mathbf{x}}$ is weighted according to their respective uncertainties.
In practice, variances are parameterised following \cite{moliner2023solving, shen2024understanding} as
$
\sigma_y^2 = \frac{\sigma(\tau)}{\omega_y \sqrt{n}} \left\| \nabla_{\mathbf{x}_\tau} \mathbf{M}(\mathbf{y} - \hat{\mathbf{x}}_0(\mathbf{x}_\tau)) \right\|_2,
$
with $\sigma_{\tilde{x}}^2$ defined analogously using a parameter $\omega_{\tilde{x}}$.
The weights $\omega_y$ and $\omega_{\tilde{x}}$ control the balance between observed and auxiliary constraints and can be set to zero to disable either contribution.

\subsection{Waveform Similarity Search} \label{sec:sim}

Inspired by \cite{perraudin_inpainting_2018}, 
we retrieve a candidate for the guide $\tilde{\mathbf{x}} \in \mathbb{R}^n$ by conducting a similarity search over a separate corpus of source audio signals, denoted by $\mathcal{S} = \{\mathbf{s}_1, \mathbf{s}_2, \dots, \mathbf{s}_N\}$, where each signal  $\mathbf{s}_i \in \mathbb{R}^{L_i}$, 
can have arbitrary length $L_i > n$. 
The goal is to identify a segment from the corpus that best matches the observed context surrounding the masked region in $\mathbf{y}$.
Unlike \cite{perraudin_inpainting_2018}, which allowed flexible segment lengths to improve matching, we search for a candidate of exactly the same length as the mask and rely on diffusion-based inference to ensure smooth transitions between the reconstructed segment and its surrounding context.


Let $\hat{t}$ denote the start of a candidate segment in $\mathbf{s}_i$, and define the segment extraction as $\mathbf{s}_i(\hat{t}) = \mathbf{s}_i[\hat{t} : \hat{t}+n-1]$. 
The optimal candidate is chosen to minimise a similarity cost:
\begin{equation}
(\mathbf{s}^*, \hat{t}^*) = \arg \min_{\mathbf{s}_i \in \mathcal{S}, \hat{t}} \mathcal{J}_i(\hat{t}), 
\quad
\tilde{\mathbf{x}} = \mathbf{s}^*(\hat{t}^*).
\end{equation}
The similarity cost $\mathcal{J}_i(\hat{t})$ compares the context surrounding the gap in $\mathbf{y}$ to the candidate segment and is defined as
\begin{equation}
\mathcal{J}_i(\hat{t}) = \sum_{k=1}^K \alpha_k \, d_k \Big( 
\Phi_k(\mathbf{W}_k  \mathbf{y}), \, 
\Phi_k(\mathbf{W}_k  \mathbf{s}_i(\hat{t}))
\Big),
\end{equation}
where $\{\Phi_k\}_{k=1}^K$ are feature mappings $\Phi_k: \mathbb{R}^{L_c} \to \mathcal{F}_k$ into separate metric spaces $(\mathcal{F}_k, d_k)$. Each $d_k$ is typically a weighted Euclidean distance, and the scalars $\alpha_k$ balance the contributions of the different feature spaces.
The weighting matrix $\mathbf{W}_k = \mathrm{diag}(\mathbf{w}_k)$ isolates the context before and after the gap, whilst providing a similarity importance weighting within the context regions. We define
$0 \leq w_{k,i} \leq 1$ for $i \in [t_s-L_c, t_s] \cup [t_e, t_e+L_c]$ and $w_{k,i}=0$ otherwise,
where $t_s$ and $t_e$ are the start and end of the masked region, and $L_c$ is the length of the context used for the similarity comparison. 

For large corpora, an exhaustive search over all possible segment start times $\hat{t}$ can be computationally expensive. 
To reduce the cost, we first perform the similarity search at a lower temporal resolution, evaluating $\mathcal{J}_i(\hat{t})$ at intervals of hop length $L_h$. 
Once a coarse optimum $\hat{t}^*$ is identified, we refine the selection in the vicinity of $\hat{t}^*$ at full resolution using a waveform-based continuity criterion to ensure smooth transitions in the chosen segment.

We define a small search window of sample offset values, $\mathcal{W} = \{-L_h/2, ..., +L_h/2\}$, around the initial estimate, $\hat{t}^*$. The offset, $o^*$, is found using the absolute value between samples at the mask's boundaries and the candidate's context,
\begin{equation}
o^* = \arg\min_{o \in \mathcal{W}} \left| \mathbf{y}_{[t_s - 1]} - \mathbf{s}_{[t^*+t_s + o]} \right| + \left| \mathbf{y}_{[t_e+1]} - \mathbf{s}_{[t^* +t_e+ o ]} \right| \ .
\end{equation}
The final starting position for the inpainting patch is then given by $t^* + o^*$, and the corresponding segment
$\tilde{\mathbf{x}} = \mathbf{s}^*_{[t^* + o^* , t^* + o^* + n-1]}$
is then extracted (with zero-padding as needed).
Through this two-stage search method, our process allows for both efficient exploration of the feature space and precise alignment of the chosen replacement segment.

\begin{figure*}[t]
    \centering
    \includegraphics[trim=7 11 6 7 , clip, width=0.85\textwidth]{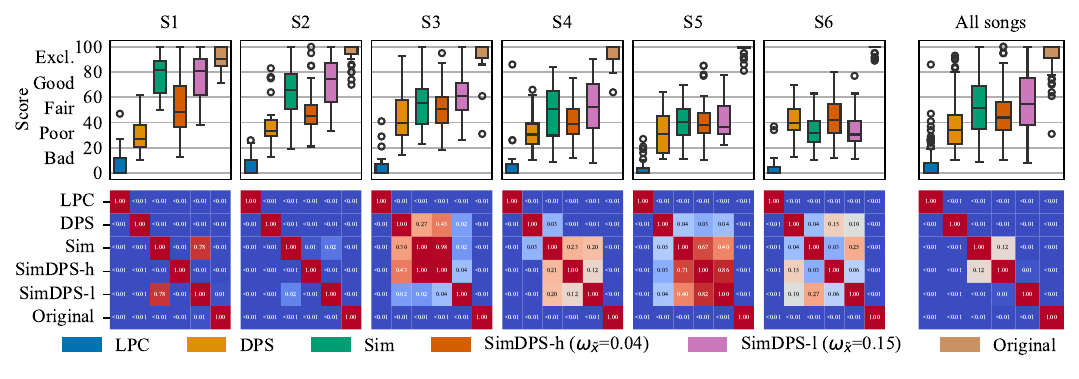}
    \vspace{-9pt}
    \caption{
(top) Boxplots of listening test scores, per example and aggregated across all songs, showing the superiority of the proposed \textit{SimDPS-l} method.  
(bottom) Tables of p-values from Wilcoxon signed-rank tests; blue cells (p$<0.05$) indicate statistically significant differences.}
    
    \label{fig:boxplots}
\end{figure*}

\vspace{-5pt}
\section{Experiments}

Although the proposed method is general, our experiments focus on the inpainting of piano recordings. 
Our experiments were conducted on 6-s signals, with gaps of 2\,s. For the similarity search, we used the remainder of each music track as the source, leveraging within-track correlations that arise from the repetitive structure of music.

\vspace{-5pt}
\subsection{Training}

For the backbone score model $s_\theta$, we use the MR-CQTdiff deep learning architecture proposed in \cite{dacosta2025mrcqtdiff}, which leverages a multi-resolution Constant-Q Transform (CQT) tailored for music signals. Since the CQT is invertible, the diffusion process is defined directly in the time domain, providing flexible, high-fidelity audio generation.
Unlike latent diffusion approaches \cite{evans_stable_2024}, which often compromise acoustic quality, we prioritise generating audio suitable for direct insertion into music tracks without noticeable artifacts.

The model is trained on the MAESTRO dataset \cite{hawthorne_enabling_2019} using mono recordings at 44.1\,kHz. Although the training segments are 6\,s long, the model being convolutional can generalize to longer sequences. Initial training was performed on the full MAESTRO train split for 180k iterations, followed by 320k iterations on recordings from 2017–2018, which have superior acoustic quality. Total training time was approximately two days on a single NVIDIA H200 GPU.

\vspace{-2pt}
\subsection{Inference}

To sample from the diffusion model, we discretise the time $\{\tau\}_{i=1}^T$ into $T=50$ steps. 
We set a maximum $\tau_T=8$ and a minimum $\tau_1=e^{-5}$, using
a logarithmic discretization schedule,
 which emphasises precision in the trajectory regions of low signal-to-noise ratio (SNR), where high-frequency detail is determined.

Although the probability flow ODE \eqref{eq:probflow} describes a deterministic trajectory, previous studies \cite{song_score-based_2021, cao_exploring_2023, karras2022elucidating} demonstrate that a stochastic solver can exhibit benefits in sample quality, effectively smoothing over errors made in prior discretisation steps.
We found that injecting stochasticity was beneficial in our experiments, thus 
 we implement the second-order stochastic sampler proposed by Karras et al.~\cite{karras2022elucidating},
 setting the stochasticity parameter $S_{\mathrm{churn}}=10$. 

We observe that the hyperparameter $\omega_{\tilde{x}}$, which controls the uncertainty of the similarity-based guidance, has a critical impact on inpainting performance.
We denote our method with low and high uncertainty guidance as SimDPS-l ($\omega_{\tilde{x}} = 0.15$) and SimDPS-h ($\omega_{\tilde{x}} = 0.04$), respectively. Setting $\omega_{\tilde{x}} = 0$ recovers the standard Diffusion Posterior Sampling (DPS) inpainting method \cite{moliner2023solving, moliner_diffusion-based_2024}. In all cases, the reconstruction weighting is fixed at $\omega_y = 0.3$.

\subsection{Similarity Search}

To perform the feature-based similarity search, the audio signals were resampled to $12$\,kHz, improving search efficiency whilst retaining perceptually-relevant frequencies.
The similarity search was conducted on a source corpus, $\mathcal{S}$, of two separated parts: the complete music track before the gap, and the complete music track after the gap.
The maximum context length, $L_c$, was set to 3\,s.

We used the STFT and Chromagram \cite{muller_music_2021} features, represented by the feature sets, $\{\Phi_1, d_1, \mathbf{w}_1, \alpha_1\}$ and $\{\Phi_2, d_2, \mathbf{w}_2, \alpha_2\}$ respectively. For both features, we used a Hann window and hop length of 256, and window length and FFT size set to 1024 for the STFT calculation.
This allowed an efficient joint frame pre-processing. Each distance metric $d_k$ is the weighted Euclidean distance, with weighting vectors $\mathbf{w}_k$ acting as ramped heterogeneous context duration terms that provide greater weighting to frames closer to the mask boundary.
The STFT features act as a short-range timbral similarity through a vector $\mathbf{w}_1$ ramped linearly to zero from 0 s to 0.75 s. In contrast, the long-range chromagram assesses the rhythmic and melodic contents across the full context duration, with $\mathbf{w}_1$ ramping linearly. The feature hyperparameters, $\alpha_1$ and $\alpha_2$, are set to $1.0$.

\section{Subjective Evaluation}

To assess the performance of the proposed method, we conducted a multi-stimulus listening test. 
The test was implemented with webMUSHRA \cite{schoeffler2018webmushra}. Each trial presented a 6-s piano excerpt with a 2-s central gap.
The test included six reconstructions of the excerpt, including the original signal (high anchor) and five methods.

We included linear predictive coding (\textit{LPC}) \cite{kauppinen_audio_2002} as a low anchor, although it cannot fill such long gaps as it only produces stationary signals.
The other compared methods (\textit{DPS}, \textit{Sim}, and \textit{SimDPS}) are as defined in the previous section, including the high and low uncertainty variants of \textit{SimDPS}.
For \textit{Sim}, adhering to the original method \cite{perraudin_inpainting_2018}, we apply a small 10-ms cross-fade at the mask boundary to avoid clicks.
Excerpts of the listening test, alongside additional examples on other types of music, can be found at our webpage\footnote{https://s-turland.github.io/SimDPS/}.

Listeners were asked to rate the plausibility of each reconstruction on a continuous scale from 0 to 100. They were encouraged, but not required, to use the full range of scores. Six different excerpts were used, each presented on two separate pages to both increase the number of ratings and assess response consistency.
A total of 17 volunteers participated, 15 of whom had prior experience with listening tests.
Their average age was 30 years old.
One participant was excluded from the analysis due to unreliable responses.

The listening test results,  shown in Fig.~\ref{fig:boxplots}, indicate a strong dependency of reconstruction quality on the individual examples, confirming the example-dependent nature of long-gap music inpainting.
Therefore, we analyse the ratings on a per-excerpt basis. Across all excerpts, the reference signals (\textit{Original}) were consistently rated as Excellent, while the anchor (\textit{LPC}) was consistently placed in the Poor range. The scores of the similarity-based condition (\textit{Sim}) provide a useful indicator of how well a musically coherent match could be found in each case. For excerpt S1, the \textit{Sim} condition reached Excellent ratings, and the low uncertainty \textit{SimDPS-l} achieved an equivalent distribution of scores, suggesting that when the retrieved match is highly coherent, our diffusion-based approach can preserve its quality.
For excerpt S2, \textit{Sim} was rated in the Good range, but \textit{SimDPS-l} obtained significantly higher ratings (p = 0.02), showing that the diffusion process can improve upon moderately fitting matches. 
In excerpts S3 and S4, where \textit{Sim} was only rated Fair, \textit{SimDPS-l} again outperformed it: significantly in S3 (p = 0.02) and marginally in S4 (p = 0.2). By contrast, for excerpts where \textit{Sim} was close to the Poor range, no consistent benefit of \textit{SimDPS-l} was observed, and both conditions yielded similar score distributions.

The unguided \textit{DPS} condition  was generally rated in the lower part of the scale, highlighting the importance of guidance for plausibility. The high uncertainty version, \textit{SimDPS-h}, often outperformed \textit{DPS} but remained below or equal to \textit{Sim} in most cases. 
An exception occurred in excerpt S6, where \textit{DPS} and \textit{SimDPS-h} were rated higher than both \textit{Sim} and \textit{SimDPS-l}, although all methods remained in the lower part of the scale. Across the full range of match quality found through the similarity search, the proposed \textit{SimDPS-l} method achieves the highest mean plausibility score.

\section{Conclusions}
\vspace{-5pt}
We proposed a music inpainting method capable of reconstructing long gaps---exceeding 0.3\,s---that combines a similarity search procedure with a generative diffusion model. This hybrid approach uses an auxiliary signal as a guide, designing a non-uniform posterior covariance matrix to direct the sample toward plausible solutions that align with the surrounding context. The method, called SimDPS, is subjectively evaluated on reconstructing piano music. Results show that SimDPS can generate samples that consistently improve on the perceptual plausibility of unguided diffusion-based inpainting, and can frequently outperform the similarity search method alone, for moderately-similar candidates.


\section{Acknowledgment}
This work was supported by Nokia Solutions and Networks. We acknowledge the computational resources from Aalto University “Science-IT” project.

\section{Compliance with Ethical Standards}

The listening test with volunteers, involving only anonymous and non-sensitive audio ratings, was conducted in accordance with the Declaration of Helsinki (1975, revised 2000).

\end{document}